# Universal inverse power-law distribution for temperature and rainfall in the UK region

A.M.Selvam[1]

B1 Aradhana, 42/2A Shivajinagar, Pune 411005, India

Email: amselvam@gmail.com

Websites: http://amselvam.webs.com; http://amselvam.tripod.com; http://www.geocities.ws/amselvam

## Abstract

Meteorological parameters, such as temperature, rainfall, pressure etc., exhibit selfsimilar space-time fractal fluctuations generic to dynamical systems in nature such as fluid flows, spread of forest fires, earthquakes, etc. The power spectra of fractal fluctuations display inverse power-law form signifying long-range correlations. The author has developed a general systems theory which predicts universal inverse power-law form incorporating the golden mean for the fractal fluctuations of all size scales, i.e., small, large and extreme values. The model predicted distribution is in close agreement with observed fractal fluctuations in the historic month-wise temperature (maximum and minimum) and rainfall in the UK region. The present study suggests that fractal fluctuations result from the superimposition of an eddy continuum fluctuations. The observed extreme values result from superimposition of maxima (or minima) of dominant eddies (waves) in the eddy continuum.

---

[1] Retired Deputy Director, Indian Institute of Tropical Meteorology, Pune 411008, India

# Universal inverse power-law distribution for temperature and rainfall in the UK region

A.M.Selvam[2]

B1 Aradhana, 42/2A Shivajinagar, Pune 411005, India

Email: amselvam@gmail.com

Websites: http://amselvam.webs.com; http://amselvam.tripod.com; http://www.geocities.ws/amselvam

**Abstract**

Meteorological parameters, such as temperature, rainfall, pressure etc., exhibit selfsimilar space-time fractal fluctuations generic to dynamical systems in nature such as fluid flows, spread of forest fires, earthquakes, etc. The power spectra of fractal fluctuations display inverse power-law form signifying long-range correlations. The author has developed a general systems theory which predicts universal inverse power-law form incorporating the golden mean for the fractal fluctuations of all size scales, i.e., small, large and extreme values. The model predicted distribution is in close agreement with observed fractal fluctuations in the historic month-wise temperature (maximum and minimum) and rainfall in the UK region. The present study suggests that fractal fluctuations result from the superimposition of an eddy continuum fluctuations. The observed extreme values result from superimposition of maxima (or minima) of dominant eddies (waves) in the eddy continuum.

## 1. Introduction

Dynamical systems such as fluid flows, heart beat patterns, spread of infectious diseases, etc., exhibit selfsimilar, i.e., a zig-zag pattern of successive increases followed by decreases of all scales identified as fractal fluctuations. Fractal fluctuations signify non-local connections, i.e., long-range correlations in space and time manifested as inverse power-law form $f^{\alpha}$ for the power spectra. Extensive studies by Lovejoy and Schertzer (2012), Bunde et al. (2013) have identified conclusively the selfsimilar fractal nature of fluctuations in meteorological parameters. The Gaussian probability distribution used widely for analysis and description of large data sets underestimates the probabilities of occurrence of extreme events such as stock market crashes, earthquakes, heavy rainfall, etc. The assumptions underlying the normal distribution such as fixed mean and standard deviation, independence of data, are not valid for real world fractal data sets exhibiting a scale-free power-law distribution with fat tails (Selvam, 2009). The author has developed a general systems theory model (Selvam 1990, 2009, 2012a, 2012b, 2013) for fractal fluctuations in dynamical systems. The model predicts universal inverse power-law form incorporating the golden mean ($\tau \approx$ 1.618) for the probability distribution of amplitudes of fractal fluctuations. The model predictions are in agreement with historic month-wise temperature (maximum and minimum) and total rainfall for four stations in the UK region for periods ranging from

---
[2] Retired Deputy Director, Indian Institute of Tropical Meteorology, Pune 411008, India

160 to 92 years. A review of 1/*f* noise in physical phenomena, general systems theory model predictions for fractal fluctuations in dynamical systems, details of data and analysis techniques used are presented in the following.

## 2. 1/*f* Noise in Physical Phenomena

Noise and fluctuations are ubiquitous to all physical phenomena and is a multidisciplinary field of study (Abbott 2001). The roots of noise research trace back to the Scottish botanist Robert Brown who observed fluctuating pollen on the surface of a film of water in 1822. The first unsolved noise problem was to find out the origin of Brownian motion (Abbott 2001).

The noise spectrum takes on an approximately 1/*f* shape at low frequencies in many real systems and is referred to as 1/*f* noise. The underlying mechanisms and the unifying features characterizing the ubiquitous occurrence of 1/*f* noise are the major unsolved problems in the area of noise and fluctuation research. The 1/*f* or "1/*f*-like" noise (referred to as power-law noise spectrum indicates that (i) high energy events tend to occur less frequently than low energy events (ii) The fluctuations are scale-free, i.e. occur on all size scales, the amplitude of fluctuations being dependent on the scale factor (space-time) alone. (iii) 1/*f* noise or power-law implies long-range (space-time) correlations signifying memory or persistence in the dynamical evolution of the fluctuations, e.g. warm to cold weather changes take place gradually. (iv) Small, large and extreme events belong to the same population, the same distribution, and reflect the same underlying mechanisms (Sornette and Ouillon 2012).

The scale-invariant character of 1/*f* noise indicates fractal or selfsimilar nature of fluctuations and is associated with nonlinear processes and may be a signature of self-organised criticality proposed in 1987 by Bak et al. (1987). Mandelbrot (1977) associated power-law distributions with fractal structures. The study of 1/*f* noise first started in electronics (Milotti 2001), with the investigation of $f^{\alpha}$ noise in vacuum tubes by Johnson (1925) and Schottky (1926).

The first quantitative discovery of power-law was Pareto's study (Pareto 1896) on the uneven distribution of incomes (Baek et al 2011; Pinto et al. 2012). Power-law distributions, encountered in a huge variety of vastly different systems must be traced to signatures of a global feature intrinsic to the dynamical processes underlying the noise/fluctuations.

The 1/*f* or "1/*f*-like" noise spectrum is ubiquitous in meteorology (Blender et al. 2011). Meteorological parameters such as temperature, wind speed etc. exhibit power-law spectrum on all space-time scales (Lovejoy and Schertzer 2012) from the surface layers to stratospheric levels and above. The classical Kolmogorov's -5/3 power-law spectrum (Kolmogorov 1941a-c) for turbulent fluctuations also belongs to 1/*f* noise category. Studies in the early 1970s by Vinnichenko and others (Vinnichenko and Dutton 1969) also showed power-law spectrum for meteorological parameters in the troposphere and higher levels.

A characteristic feature of climatic records is their pronounced variability. The spectral analysis of continuous climatic time series normally reveals a continuous variance distribution encompassing all resolvable frequencies, with higher variance levels at lower frequencies or "red noise" (Hasselmann 1976) typical of 1/*f* noise. Understanding the natural variability of climate is one of the most important tasks facing climatologists (Pelletier 1997).

Fraedrich (2002, 2009) gives a brief summary of studies relating to 1/*f* noise in climate data. In the mid-1970s the Brownian motion entered climate research as an analog for the earth's climate fluctuations (Hasselmann 1976) which led to an intensive red noise search in data. At the same time observations and modeling of flicker noise and other power-law scaling regimes emerged (for example Voss and Clarke 1976, Vliet et al. 1980) with concepts as close to the climate systems energy balance as the Brownian motion analog. Since then power-law power spectra different from Brownian motion has been identified in observed records and model simulations of the climate system. While most of these studies (for a review see Pelletier and Turcotte 1999) are guided by self-affine scaling laws governing the dynamics of a nonlinear system, the associated long-range memory or correlation aspect has been emphasized only recently (Koscielny-Bunde et al. 1998, Talkner and Weber 2000 analysing observed temperatures). The analyses suggest that the near surface temperature fluctuations are governed by a universal scaling behavior showing long-term memory correlations up to at least 30 years. Huybers and Curry (2006) demonstrate that climate variability exists at all timescales with climate processes being intimately coupled.

## 3. General Systems Theory for Fractal Fluctuations

Power (variance) spectra of fractal fluctuations exhibit inverse power-law form $f^{\alpha}$ where *f* is the frequency (or wavelength of the eddies) and α the exponent indicating (i) selfsimilar fractal fluctuations result from the coexistence of a continuum of eddies (waves) (ii) fractal fluctuations exhibit long-range space-time correlations since the amplitudes of larger and smaller size eddies are related to each other by the scale factor α alone independent of other characteristics of the eddies.

The general systems theory model (Selvam, 1990, 2007, 2012a, 2012b, 2013) is based on the above observational fact that fractal fluctuations signify an underlying eddy continuum. The model is based on the simple concept that large eddies result from successive space-time integration of enclosed small-scale fluctuations (eddies) analogous to Townsend's (1956) concept that large eddies are envelopes enclosing smaller scale eddies. The model predictions are as follows.

1. Fractal fluctuations result from the superimposition of eddy fluctuations of an eddy continuum, the component eddy circulations tracing the quasiperiodic Penrose tiling pattern.

2. The eddy continuum is generated starting from unit primary eddy (radius *r*). Starting with unit primary eddy, successive stages of large eddy growth is associated with scale ratio $z =$ to 1, 2, 3, etc. The primary eddy growth region is $z = 0$ to 1.

3. The amplitude and variance of fractal fluctuations (space/time) exhibit the same (identical) inverse power law distribution *P* with respect to normalized standard deviation $t$ equal to mean/standard deviation. Fractal fluctuations therefore signify quantumlike chaos since the property that the additive amplitudes of eddies when squared represent the probability densities is exhibited by the subatomic dynamics of quantum systems such as the electron or photon.

4. The probability distribution $P = \tau^{-4t}$ holds for the range of normalized deviation $t$ values $t \geq 1$ and $t \leq -1$.

5. The primary eddy growth corresponds to normalised deviation $t$ ranging from -1 to +1. In this region the probability $P$ is shown to be equal to $P = \tau^{-4k}$ where $k = \sqrt{\dfrac{\pi}{2z}}$ is the steady state fractional volume dilution $k$ of the growing primary eddy by internal smaller scale eddy mixing (Selvam 2013).

6. The model predicted universal inverse power-law distribution is very close to the statistical normal distribution for normalized deviation $t$ values less than 2 and exhibits a long fat tail for $t$ values more than 2, i.e., extreme events have a higher probability of occurrence than that predicted by statistical normal distribution as found in practice. The statistical normal distribution and the model predicted universal inverse power-law distribution are shown in Fig.1 (Selvam, 2013, http://arxiv.org/abs/1111.3132).

## 4. Data

Month-wise climate data of temperature (maximum and minimum) and rainfall for four stations (i) Oxford (ii) Armagh (iii) Durham (iv) Stornoway (Table 1) in the UK region were obtained from http://www.metoffice.gov.uk/climate/uk/stationdata.

Table 1.

| Station | Maxm Temperature | Minm Temperature | Rainfall |
|---|---|---|---|
| | Duration (years) | | |
| Oxford 45.09E 20.72N, 63 metres amsl | 160 (1853 – 2012) | 160 (1853 – 2012) | 160 (1853 – 2012) |
| Armagh 28.78E 34.58N (Irish Grid), 62 metres amsl | 147 (1866 - 2012) | 147 (1866 - 2012) | 160 (1853 - 2012) |
| Durham 42.67E 54.15N, 102 metres amsl | 133 (1880 - 2012) | 133 (1880 - 2012) | 133 (1880 - 2012) |
| Stornoway 14.64E 93.32N, 15 metres amsl | 139 (1874 - 2012) | 92 (1921 - 2012) | 139 (1874 - 2012) |

## 4.1 Analyses and results

Each data set was represented as the frequency of occurrence $f(i)$ in a suitable number $n$ of class intervals $x(i)$, $i=1$, $n$ covering the range of values from *minimum* to the *maximum* in the data set. The class interval $x(i)$ represents dataset values in the range $x(i) \pm \Delta x$, where $\Delta x$ is a constant. The average *av* and standard deviation *sd* for the data set is computed as

$$av = \frac{\sum_{1}^{n}[x(i) \times f(i)]}{\sum_{1}^{n} f(i)}$$

$$sd = \frac{\sum_{1}^{n}\{[x(i) - av]^2 \times f(i)\}}{\sum_{1}^{n} f(i)}$$

The *normalized deviation t* values for class intervals $t(i)$ were then computed as

$$t(i) = \frac{x(i) - av}{sd}$$

The cumulative percentage probabilities of occurrence $cmax(i)$ and $cmin(i)$ were then computed starting respectively from the maximum ($i=n$) and minimum ($i=1$) class interval values as follows.

$$cmax(i) = \frac{\sum_{n}^{i}[x(i) \times f(i)]}{\sum_{1}^{n}[x(i) \times f(i)]} \times 100.0$$

$$cmin(i) = \frac{\sum_{1}^{i}[x(i) \times f(i)]}{\sum_{1}^{n}[x(i) \times f(i)]} \times 100.0$$

The month-wise (January to December) average and standard deviation of cumulative percentage probability values $cmax(i)$ and $cmin(i)$ for temperature and rainfall were computed for the four UK stations and plotted with respect to corresponding *normalized deviation* $t(i)$ values with logarithmic scale for the probability axis (Figs. 2 and 3) along with model predicted universal inverse power-law distribution. There is a close correspondence between model predicted and observed probability distributions of amplitudes of fractal fluctuations of all size scales (small, large and extreme) in UK region monthly temperature (maximum and minimum) and monthly total rainfall.

### Average probability distribution of fractal fluctuations monthly mean temperature (maxm and minm) and rainfall

Oxford for 160 years (1853 -2012)

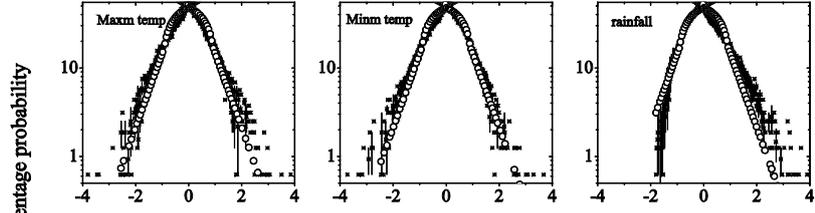

Armagh temperature for 147 years (1853 -2012) and rainfall for 160 years (1853 - 2012)

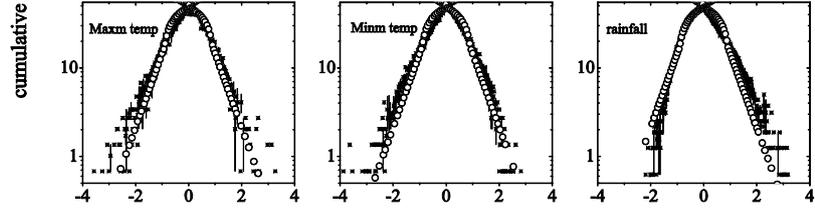

Fig. 2: The month-wise (January to December) average and standard deviation of cumulative percentage probability values for Oxford and Armagh plotted with respect to corresponding *normalized deviation t* values with logarithmic scale for the probability axis along with model predicted universal inverse power-law distribution.

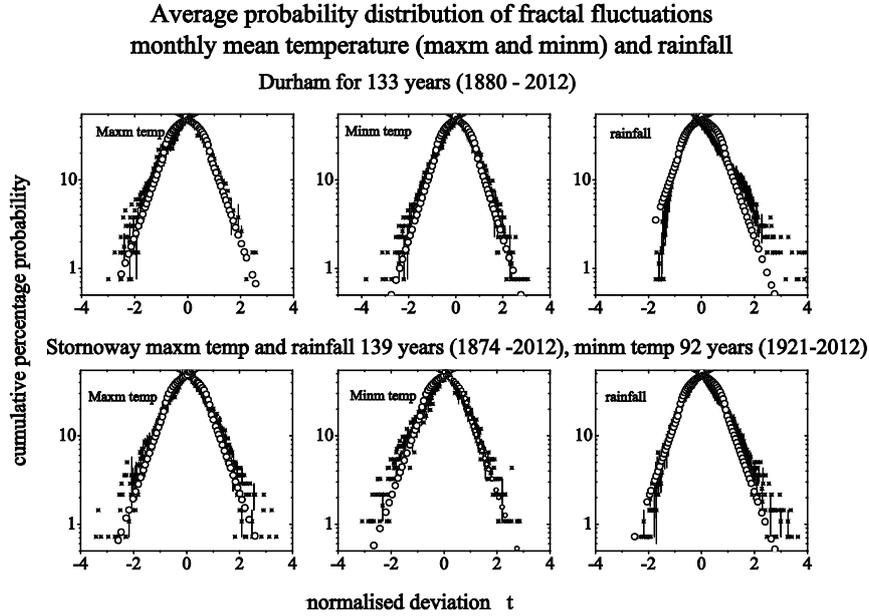

Fig. 3: The month-wise (January to December) average and standard deviation of cumulative percentage probability values for Durham and Stornoway plotted with respect to corresponding *normalized deviation t* values with logarithmic scale for the probability axis along with model predicted universal inverse power-law distribution.

## 5. Discussion

The probability distribution *P* of amplitudes of fractal fluctuations in UK region month-wise (January – December) temperature (maximum and minimum) and total rainfall for fluctuations of all size scales (small, large and extreme) closely follows the general systems theory model predicted universal inverse power-law distribution $P = \tau^{-4t}$ where $\tau$ is the golden mean ($\approx 1.618$) and *t* the normalized deviation equal to mean/standard deviation. The model predicted distribution is close to the observed distribution particularly for the normalized deviation *t* values greater than 2 which correspond to extreme events with higher probability of occurrence than that predicted by the statistical normal distribution.

Inverse power-law distribution for fractal fluctuations implies long-range space-time correlations manifested as memory or persistence in the space-time variability of the meteorological parameter such as rainfall, temperature, etc. Kantelhardt et al. (2006) state that the persistence analysis of river flows and precipitation has been initiated, about half a century ago, by H. E. Hurst, who found that runoff records from various rivers exhibit ''long-range statistical dependencies'' (Hurst 1951). Later, similar long-term correlated fluctuation behavior has also been reported for many other geophysical records including temperature and precipitation data (Kantelhardt et al. 2006). Characterizing and understanding the persistence of wet and dry conditions in the distant past gives new perspectives on contemporary climate change and its causes (Bunde et al. 2013).

## 6. Conclusion

A general systems theory model (Selvam 1990, 2013) predicts universal inverse power-law form incorporating the golden mean for the fractal fluctuations. The model predicted distribution is in close agreement with observed fractal fluctuations of all size scales (small, large and extreme values) in the historic month-wise temperature (maximum and minimum) and total rainfall for the four stations in the UK region, Oxford, Armagh, Durham and Stornoway for the data periods ranging from 92 years to 160 years.

The present study suggests that fractal fluctuations result from the superimposition of an eddy continuum fluctuations and as such the observed extreme values may result from superimposition of maxima (or minima) of dominant eddies (waves) in the eddy continuum.

## Acknowledgement

The author expresses her gratitude to the UK Meteorological Office for providing the temperature and rainfall data used in this study. This research work was carried out after retirement by the author alone without any funding. The author is grateful to Dr. A. S. R. Murty for encouragement during the course of the study.